\documentclass[aps,prl,twocolumn,showpacs,a4paper]{revtex4}
\usepackage{graphicx}

\newcommand{\br}{{\mathbf r}}
\newcommand{\e} {{\mathrm e}}

\begin{document}

\title{Correlated few-particle states in artificial bipolar molecule}
\author{Egidijus Anisimovas}
\email{egidijus@uia.ua.ac.be}
\author{F. M. Peeters}
\email{peeters@uia.ua.ac.be}
\affiliation{Departement Natuurkunde, Universiteit Antwerpen (UIA),
B-2610 Antwerpen, Belgium} 

\date{22 January 2002}

\begin{abstract}
\vspace{3mm}
We investigate the ground and excited states of a bipolar artificial
molecule composed of two vertically coupled quantum dots containing 
different type of carriers -- electrons and holes -- in equilibrium. 
The approach based on exact diagonalization is used and reveals an 
intricate pattern of ground-state angular momentum switching and a 
rearrangement of approximate single-particle levels as a function of 
the inter-dot coupling strength.
\end{abstract}


\maketitle



In recent decades, artificially created zero-dimensional systems ---
the so-called quantum dots or artificial atoms --- have been a major
focus of active research in condensed matter physics \cite{jacak98}.
Most of the interest arises from our ability to control the addition
or extraction of even a single particle \cite{tarucha96,hawr00} as 
well as from the fact that these artificial structures often offer a 
rich set of novel features unrivaled by those of their natural 
counterparts. For example, while the nature of atomic spectra is 
mostly governed by the quantization of motion of a single particle 
in the field of the nucleus, in quantum dots the crucial role of the
quantizing magnetic field and strong Coulomb interaction between 
the electrons leads to the formation of intriguing `magic' states
\cite{jacak98,magic} related to the compact states of non-interacting
composite fermions \cite{jacak98,composite}.

The ability to create and study double-layer systems has added a new 
dimension to the physics of quantum dots. Two vertically stacked quantum 
dots make an example of an artificial molecule. The observation of a 
novel incompressible state induced by the inter-layer coupling 
\cite{murphy94} motivated a study of correlated few-particle states in 
vertical double-dot systems \cite{palacios95}. In this work, the inter-dot 
interactions were analysed in terms of an isospin-$1/2$, and a new class 
of incompressible ground states was found.

In the present paper, we examine the correlated few-particle states of
an artificial {\em bipolar} molecule consisting of two vertically stacked
quantum dots populated by carriers of different kinds -- electrons and
holes, respectively. Such a system could be structured, for example, in
bilayer-bipolar heterostructures containing electron and hole layers 
{\em in equilibrium}. These structures have been realized in the crossed
gap InAs/GaSb system \cite{inas} as well as in biased GaAs/AlGaAs 
heterostructure \cite{gaas,sivan92} where electron and hole layers
form on the opposite sides of an AlGaAs barrier. Most of the interest 
in such systems stems from the possibility (at least, in principle) of the 
formation of Bose-Einstein condensate of indirect excitons \cite{lozovik99}.

While the superfluid state has not been demonstrated so far, a number 
of other interesting effects of electron-hole coupling and hybridization
have been predicted and observed. For example, the Coulomb-drag experiments
\cite{sivan92,tso93} revealed the inter-layer momentum transfer 
rates up to an order of magnitude larger than in similar electron-electron 
systems, and magnetotransport calculations \cite{naveh01} established the
presence of coupling-dependent current that remains large even when the
Fermi energy lies in the hybridization gap.

In the present paper, we consider two
vertically coupled infinitesimally thin circularly-symmetric quantum 
dots placed into a strong perpendicular magnetic field. The confining 
potentials of the 
dots are assumed to be of a simple parabolic shape, and we choose the 
base frequencies of the two dots to be inversely proportional to the 
effective masses of the respective carriers, i.e. 
$\omega_e m_e = \omega_h m_h$. This choice implies the equality of 
the quantum dot oscillator lengths $l_0 = \sqrt{\hbar/m_a\omega_a}$ 
for $a = e, h$ (a plausible assumption) as well as the equality of 
renormalized oscillator lengths in magnetic fields of arbitrary
strength since $l^{-4} = l_0^{-4} + (1/4) l_c^{-4}$ and the cyclotron 
radius $l_c = \sqrt{\hbar c/eB}$ does not depend on the material 
parameters. The above constraint also considerably simplifies the 
calculation of Coulomb matrix elements between the electron and hole
states (see below).

In the limit of strong magnetic fields, when the cyclotron frequencies 
of the two types of carriers $\omega_{c(e)}$ and $\omega_{c(h)}$ exceed 
the respective base dot frequencies (in practice, $B$ $>$ 2--3 T) we can 
restrict the single-particle bases in the dots to the essentially 
one-dimensional sets of the lowest-Landau-level states. 
The energies of these states are spaced 
equidistantly by $\varepsilon_a = \hbar \omega_a^2/\omega_{c(a)}$ with 
$a = e,h$. We note that due to the inverse proportionality of the 
cyclotron frequency to the effective mass we have 
$\varepsilon_e/\varepsilon_h = m_h/m_e$. 

These states are labelled by a single index $n = 0,1,2 \ldots$ 
\begin{equation}
\label{eq_basis}
  \psi_n(r,\theta) = \frac{1}{\sqrt{\pi n!}} \,\e^{\mp i n \theta}
  \frac{r^n}{l^{n+1}} \e^{-r^2/2l^2},
\end{equation}
where the upper sign is to be used for electrons and the lower one for 
holes. That is, the electrons can only have negative or zero angular 
momenta and vice versa. Moreover, since the particles in strong magnetic 
fields are spin-polarized we can neglect the presence of the spin degree 
of freedom and also drop the constant Zeeman energy.

For a many particle system, a convenient basis is given by the set of 
all possible Slater determinants (equivalently, ordered strings of the 
associated creation operators) composed of the functions (\ref{eq_basis}). 
Thus we will denote possible configurations of the molecule by enumerating
the single-particle quantum numbers $n_i$ in ascending order, first for
electrons and then, separated by a semicolon, for holes. E.g. the ground 
state of three non-interacting electrons and holes is given by
\{012;012\}. Due to the cylindrical symmetry of the molecule 
the many-body states of different total angular momenta $M_t$ will not 
be mixed, thus we will be able to perform diagonalizations separately 
in subspaces of a given $M_t$. As a matter of fact, in our basis the 
single-body part of the Hamiltonian is diagonal and the remaining task
is the calculation and diagonalization of the Coulomb interactions.

Before proceeding any further, it is convenient to introduce handy
dimensionless units thus also clarifying the relevant energy and
length scales of the problem. From now on we will measure all the 
energies in $\varepsilon_e$ and the lengths in $l$. Then the spacing 
of the electronic single-particle is set equal to $1$, and the spacing 
of the hole levels is $m_e/m_h$. The interaction part of the 
Hamiltonian becomes
\begin{eqnarray}
\label{eq_int}
  V &=& \lambda \bigg\{ \sum_{i>j}^{N_e} \frac{1}{|\br_i - \br_j|}
    + \sum_{i>j}^{N_h} \frac{1}{|\br_i - \br_j|} \nonumber\\ 
    & & - \sum_{i}^{N_e} \sum_{j}^{N_h} \frac{1}{|\br_i - \br_j + 
    \vec\Delta|} \bigg\}.
\end{eqnarray}
Here, $\vec\Delta$ represents the vertical separation between the two
dots, and 
\begin{equation}
\label{}
  \lambda = \frac{l_0^4}{a_B^{*} l_c^2 l} \approx 
    \frac{l_0^4}{\sqrt{2}a_B^{*} l_c^3} \sim B^{3/2}
\end{equation}
expresses the relative strength of inter-particle interactions in
terms of a ratio of the characteristic lengths, $a_B^{*}$ being the
effective {\em electronic} Bohr radius.

We calculate the matrix elements of the Coulomb interaction between
various many-body states by decomposing them into a sum 
of two-particle Coulomb integrals
\begin{equation}
\label{eq_coul}
  J = \int\!\!\! d^2 r_1 \int\!\!\! d^2 r_2 \,
    \psi_{n_1}^{*} (\br_1) \psi_{n_2}^{*} (\br_2) 
    \,v\, \psi_{n_3} (\br_1) \psi_{n_4} (\br_2)
\end{equation}
with $v = |\br_1 - \br_2|^{-1}$ if both particles belong to the same 
layer, and $v = |\br_1 - \br_2 + \vec\Delta|^{-1}$ otherwise. In the 
former case the integral is easily evaluated analytically using the 
complex-variable technique due to Girvin and Jach \cite{girvin83} to 
separate the CM and relative coordinates (see formula (7) in their paper).
When the particles $\br_1$ and $\br_2$ belong to different layers the 
only modification is the replacement $\rho^{-1} \to (\rho^2 + \Delta^2)^{-1/2}$
in the integral over the radial part of the relative coordinate $\rho$
which has to be computed numerically.


We numerically calculated energy levels and 
many-particle states of neutral systems consisting of equal numbers 
($1$, $2$ and $3$) of electrons and holes in the dots. However, in order
to keep the presentation concise, we will concentrate 
on the most illustrative results obtained for a six-particle artificial 
molecule. Systems involving smaller number of particles show qualitatively 
similar behaviour.

\begin{figure}
\vspace{3mm}
\includegraphics[width=0.45\textwidth]{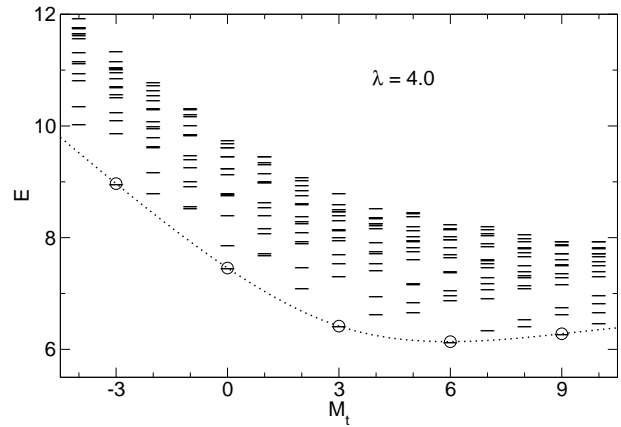}
\caption{Angular-momentum resolved spectrum of an artificial molecule
in the weak-coupling limit. The stable low-energy `magic' states are
marked by open dots and joined by a dashed line.}
\label{fig_yrast}
\end{figure}

{\it Weakly correlated dots.}
When the distance between the two dots is relatively large --- such 
as $\Delta = 3.3$ (in the units $l$) to be used in the present example 
--- the inter-dot interactions between the particles are rather weak 
compared to the strong Coulomb repulsion within the individual dots 
which is decisive in the determination of the overall behaviour of 
the molecule. In Fig.\ \ref{fig_yrast} we plot the angular-momentum 
resolved spectrum of the system showing $15$ 
lowest states for each value of $M_t$ at $\lambda = 4.0$. One sees 
that for the values of $M_t$ equal to multiples of $3$ the lowest-energy 
states (marked by the open circles) are typically considerably lower 
than those corresponding to the neighbouring values of the angular 
momentum. This is a consequence of the fact that the individual dots
`prefer' to be in the well-known `magic' states. For three particles 
(electrons or holes) the predominant single-particle configurations
in these stable many-body states are \{012\}, \{123\}, \{234\}, etc. It will 
be instructive to compare these configurations with their counterparts 
observed in strongly coupled dots below.

Since the effective mass of the holes is considerably larger than
that of electrons (we use $m_h = 6.7 m_e$) the Coulomb repulsion
promotes them to states of higher angular momentum more easily.
In Fig.\ \ref{fig_yrast} we see that the ground state has 
$M_t = 6$, and the dashed lowest-energy curve rises much more sharply
in the direction of negative angular momenta.

\begin{figure}
\vspace{3mm}
\includegraphics[width=0.45\textwidth]{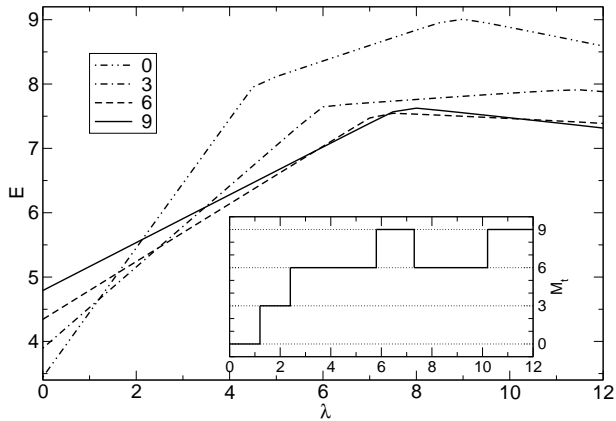}
\caption{The dependence of relative ground-state energies with $M_t = 0,3,6,9$ 
on the interaction strength $\lambda$. The inset shows the switching of the
angular momentum of the absolute ground-state.}
\label{fig_weak}
\end{figure}

While the angular momentum of the ground state is generally expected to 
grow with increasing Coulomb repulsion, this growth is not monotonous. 
The inset of Fig.\ \ref{fig_weak} shows the evolution of the ground-state 
$M_t$ versus $\lambda$. Notice that the angular momentum reaches the value 
$M_t = 9$ in three monotonous steps $\Delta M_t = 3$, and then switches 
back and forth to the value $M_t = 6$. To understand this kind of behaviour 
we also investigated the evolution of the relative ground-states (i.e. 
lowest-energy states of a given $M_t$) of the involved angular momenta in
Fig.\ \ref{fig_weak}. We see that the $\lambda$-dependences of these energies 
typically consist of segments of nearly straight lines connected by abrupt 
breaks associated with the change of the nature of the states. For example, 
we observe two breaks in the $M_t = 0$ curve at 
$\lambda \approx 4$ and $\lambda \approx 9$. At $\lambda \approx 4$ the 
molecule minimizes its energy by switching from the dominant configuration 
\{012;012\} into \{123;123\} followed by the changeover into \{234;234\} 
at $\lambda \approx 9$.
The curves corresponding to higher values of $M_t$ also show 
similar behaviour which in this particular case results in a close competition 
between the $M_t = 6$ and $M_t = 9$ lowest-energy states and switching
of the absolute ground-state in a rather extended range of $\lambda$ values.

\begin{figure}
\vspace{3mm}
\includegraphics[width=0.45\textwidth]{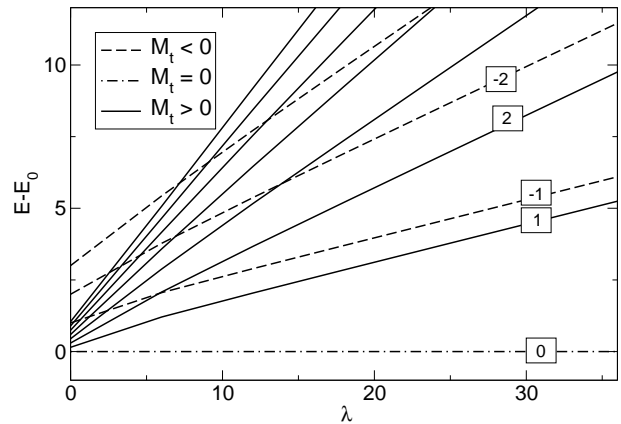}
\caption{The evolution of the relative ground-states with $-3 \le M_t \le 7$
versus the coupling strength $\lambda$. The ground state energy is subtracted
from all the curves.}
\label{fig_strong}
\end{figure}

{\it Strongly correlated dots.}
Bringing the two dots closer (we will consider the case of $\Delta = 0.2$) 
results in an entirely different character of their energy spectra. The Coulomb 
attraction between particles of different kind wins over the intra-dot 
repulsion and leads to the formation of rather strongly bound and weakly 
interacting excitons. The energies
of the relative ground-states of given angular momenta decrease rapidly and 
monotonously with increasing $\lambda$ and no longer contain abrupt breaks. 
The absolute ground state always has the total angular momentum $M_t = 0$.
For the clarity of presentation, in Fig.\ \ref{fig_strong} we plot the 
{\em differences} of the lowest-energy states of $M_t = -3 \ldots 7$ and the 
absolute ground state ($M_t = 0$). In this figure we see a clear rearrangement 
of the structure of the energy levels from the weakly interacting limit to the 
strongly-interacting limit (high $\lambda$). At $\lambda \to 0$ the levels 
corresponding to the positive momenta are spaced more closely that the 
negative-momentum levels due to the higher hole mass. For sufficiently 
high $\lambda$ values the energy levels regroup so that the levels 
corresponding to the angular momenta $M_t$ and $-M_t$ form closely spaced pairs. 

A very interesting feature of the correlated electron and hole states in 
this regime can be observed by looking more closely at the dominant single 
particle configurations of the many-body states. In the table \ref{tab} 
we compare the configurations corresponding to the relative 
ground states for $M_t = 0 \ldots 7$ in the weakly and strongly coupled 
dots. Since we are looking at the positive momentum states the electronic
subsystem is in the energetically most favourable state \{012\} of the
total momentum $M_e = -3$ while the holes contribute the angular
momentum $M_h = M_t + 3$. However, we see that the pattern of filling of
the hole states is very distinct in the two regimes. In the weak coupling
limit, the holes favour the `traditional magic' filling scheme in which 
they occupy a compact set of adjacent angular momentum orbitals, such as
\{123\} in the case of $M_h = 6$. In contrast, when the holes are in the
$M_h = 6$ states in the strong coupling limit, the configuration \{123\}
becomes the least favoured one. Instead, the dominant configuration is
\{015\} in which two of the holes occupy the orbitals of the lowest 
possible angular momentum while the third one is placed into a remote
orbital $n = 5$.

\begin{table}
\caption{\label{tab}The predominant single-particle configurations of the 
relative ground states.}
\begin{ruledtabular}
\begin{tabular}{ccc}
$M_t$ & Weak inter-dot coupling & Strong inter-dot coupling \\
\hline
0 & \{012;012\} & \{012;012\} \\
1 & \{012;013\} & \{012;013\} \\
2 & \{012;023\} & \{012;014\} \\
3 & \{012;123\} & \{012;015\} \\
4 & \{012;124\} & \{012;016\} \\
5 & \{012;134\} & \{012;017\} \\
6 & \{012;234\} & \{012;018\} \\
7 & \{012;235\} & \{012;019\} \\
\end{tabular}
\end{ruledtabular}
\end{table}

\begin{figure}
\vspace{3mm}
\includegraphics[width=0.45\textwidth]{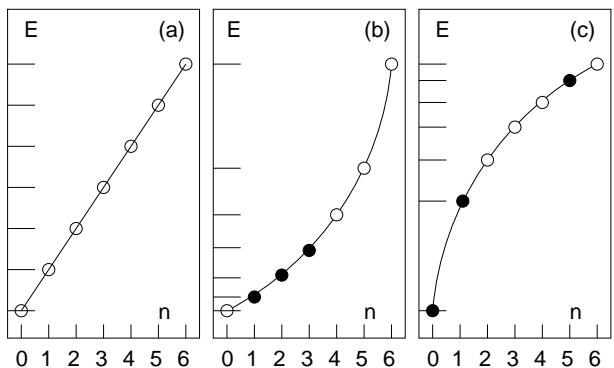}
\caption{A schematic plot of single-particle energies versus angular
momentum. Panel (a) displays the case of equidistant level spacing, 
(b) corresponds to the case of energy levels condensing downwards,
and (c) represents the case of upwards-condensing energy levels. 
Closed dots in panels (b) and (c) depict the lowest-energy 
configurations of total momentum $M_h = 6$.}
\label{fig_scheme}
\end{figure}

This new type of systematics of the single-particle configurations
can be understood by realising that singling out one particle into
a remote high-$n$ orbital becomes energetically favourable in the
case when the spacing between energy levels is decreasing with
increasing $n$, while in the opposite case the preferred configuration
is the compact one. We illustrate this point in Fig.\ \ref{fig_scheme}.
Panel (a) shows the dependence of the single-particle energy on its angular 
momentum for {\em non-interacting} particles in a quantum dot. Since the 
energy levels are equidistant, all three ways of accommodating three 
particles of the total momentum $M_h = 6$ --- \{015\}, \{024\}
and \{123\} --- have equal total energies. However, if the energy levels 
are condensing downwards [panel (b)], the preferred configuration is 
the compact \{123\} one as we observe in the weak-coupling limit.
On the other hand, if the energy levels are condensing upwards, as 
shown in panel (c), the favoured configuration is \{015\} thus 
reproducing our result in the strong-coupling limit.


In conclusion, we have investigated the many-body states of a bipolar
artificial molecule consisting of two vertically aligned quantum dots
in two distinct regimes. In the weak-coupling limit we found an 
intricate pattern of switching of the ground state angular momentum
as a function of a perpendicular magnetic field, while in the opposite 
strong-coupling limit we observe the formation of a new class of 
energetically favourable states.

This work is supported by the Flemish Science Foundation (FWO-Vl) and 
the University of Antwerp through a VIS-project. EA is supported 
by a Marie Curie Fellowship of the EU.


%



\begin{thebibliography}{99}

\bibitem{jacak98}
  L. Jacak, P. Hawrylak, and A. W\'{o}js,
  {\it Quantum Dots}, (Springer, Berlin, 1998).

\bibitem{tarucha96}
  S. Tarucha {\it et al.},
  \prl {\bf 77}, 3613 (1996).

\bibitem{hawr00}
  P. Hawrylak, G. A. Narvaez, M. Bayer, and A. Forchel,
  \prl {\bf 85}, 389 (2000).

\bibitem{magic}
  P. A. Maksym and T. Chakraborty,
    \prl {\bf 65}, 108 (1990); 
  R. B. Laughlin,
    \prb {\bf 27}, 3383 (1983).

\bibitem{composite}
  J. K. Jain and T. Kawamura,
     Europhys.\ Lett.\ {\bf 29}, 321 (1995);
  R. K. Kamilla and J. K. Jain,
     \prb {\bf 52}, 2798 (1995).

\bibitem{murphy94}
  S. Q. Murphy {\it et al.},
  \prl {\bf 72}, 728 (1994).

\bibitem{palacios95}
  J. J. Palacios and P. Hawrylak,
    \prb {\bf 51}, 1769 (1995);
  see also
  B. Partoens and F. M. Peeters,
    \prl {\bf 84}, 4433 (2000).

\bibitem{inas}
  See, for example, 
  T. P. Marlow {\it et al.}, \prl {\bf 82}, 2362 (1999);
  J. Kono {\it et al.}, \prb {\bf 55}, 1617 (1997);
  M. Altarelli, \prb {\bf 28}, 842 (1983) 
  and references therein.

\bibitem{gaas}
  Yu. E. Lozovik and V. I. Yudson,
    Pis'ma Zh.\ Eksp.\ Teor.\ Fiz.\ {\bf 22}, 556 (1975)
    [JETP Lett.\ {\bf 22}, 274 (1975)].
  
\bibitem{sivan92}
  U. Sivan, P. M. Solomon, and H. Shtrikman,
  \prl {\bf 68}, 1196 (1992).

\bibitem{lozovik99}
  Yu. E. Lozovik, O. L. Berman, and V. G. Tsvetus,
  \prb {\bf 59}, 5627 (1999).

\bibitem{tso93}
  H. C. Tso, P. Vasilopoulos, and F. M. Peeters,
    \prl {\bf 70}, 2146 (1993);
  L. \'{S}wierkowski, J. Szyma\'{n}ski, and Z. W. Gortel,
    \prl {\bf 74}, 3245 (1995).

\bibitem{naveh01}
  Y. Naveh and B. Laikhtman,
  Europhys.\ Lett.\ {\bf 55}, 545 (2001).

\bibitem{girvin83}
  S. M. Girvin and T. Jach,
  \prb {\bf 28}, 4506 (1983).

\end{thebibliography}
\end{document}